# Temperature- and field-driven spin reorientations in triple-layer ruthenate $Sr_4Ru_3O_{10}$

M. Zhu[1], P. G. Li[2], Y. Wang[2], H. B. Cao[3], W. Tian[3], H. D. Zhang[1], B. D. Phelan[1], Z. Q. Mao[2], and X. Ke[1*]

[1]*Department of Physics and Astronomy, Michigan State University, East Lansing, Michigan 48824, USA*

[2]*Department of Physics and Engineering Physics, Tulane University, New Orleans, Louisiana 70118, USA*

[3]*Neutron Scattering Division, Oak Ridge National Laboratory, Oak Ridge, Tennessee 37831, USA*

[*] Corresponding author: ke@pa.msu.edu

$Sr_4Ru_3O_{10}$, the $n = 3$ member of the Ruddlesden-Popper type ruthenate $Sr_{n+1}Ru_nO_{3n+1}$, is known to exhibit a peculiar metamagnetic transition in an in-plane magnetic field. However, the nature of both the temperature- and field-dependent phase transitions remains as a topic of debate. Here, we have investigated the magnetic transitions of $Sr_4Ru_3O_{10}$ via single-crystal neutron diffraction measurements. At zero field, we find that the system undergoes a ferromagnetic transition with both in-plane and out-of-plane magnetic components at $T_c \approx 100$ K. Below $T^* = 50$ K, the magnetic moments incline continuously toward the out-of-plane direction. At $T = 1.5$ K, where the spins are nearly aligned along the $c$ axis, a spin reorientation occurs above a critical field $B_c$, giving rise to a spin component perpendicular to the plane defined by the field direction and the $c$ axis. We suggest that both the temperature- and field-driven spin reorientations are associated with a change in the magnetocrystalline anisotropy, which is strongly coupled to the lattice degrees of freedom. This study elucidates the long-standing puzzles on the zero-field magnetic orders of $Sr_4Ru_3O_{10}$ and provides new insights into the nature of the field-induced metamagnetic transition.

**Introduction**

Ruddlesden-Popper type perovskite ruthenates $(Sr,Ca)_{n+1}Ru_nO_{3n+1}$ are prototypical $4d$ transition-metal oxides which exhibit a variety of intriguing phenomena. The magnetic and electronic properties of these materials are sensitively dependent on the layer number $n$ and the structural distortions induced by substituting Ca for Sr. For instance, the single-layer $Sr_2RuO_4$ ($n = 1$) shows an unconventional superconducting state [1], whereas the ground state of the double-layer $Sr_3Ru_2O_7$ ($n = 2$) is a Fermi liquid and is close to a ferromagnetic instability [2]. The three-dimensional $SrRuO_3$ ($n = \infty$) is a ferromagnetic metal with a Curie temperature $T_c = 160$ K [3,4]. On the other hand, $Ca_2RuO_4$ ($n = 1$) is an antiferromagnetic Mott insulator with a Neel temperature $T_N \sim 110$ K [5], while $Ca_3Ru_2O_7$ ($n = 2$) exhibits a quasi-two-dimensional metallic behavior and becomes antiferromagnetic below $T_N \sim 56$ K [6,7], and $CaRuO_3$ is a paramagnetic metal [8,9]. To date, compared to the intense efforts invested on the single-, double-layer, and three-dimensional ($n = 1, 2$, and $\infty$) ruthenates, there have been much fewer studies on the triple-layer ($n = 3$) compounds.

$Sr_4Ru_3O_{10}$ crystallizes in an orthorhombic space group *Pbam* [10], as shown in Fig. 1(a), which displays interesting but perplexing magnetic properties. From the magnetic susceptibility measurements, it has been reported that $Sr_4Ru_3O_{10}$ undergoes a paramagnetic-ferromagnetic transition at $T_c = 105$ K, below which the easy axis is along the $c$ direction [10]. Intriguingly, while an additional transition has been found in the magnetic susceptibility at $T^* = 50$ K [10,11], no anomaly is revealed in the specific heat measurements [12]. Although several distinct scenarios have been proposed to account for the feature at $T^*$, its intrinsic character remains an open question. On the one hand, below $T^*$ a canted antiferromagnetic structure with a ferromagnetic component along the $c$ axis and an antiferromagnetic component in the $ab$ plane has been proposed [11,13]. Nevertheless, no evidence of antiferromagnetism has been observed via neutron diffraction experiments [14,15]. On the other hand, previous neutron diffraction measurements have yielded contradictory results regarding the direction of the ferromagnetic moments in $Sr_4Ru_3O_{10}$ [14,15]. It is initially argued that spins are aligned in the *ab* plane based on the observation of (0 0 *L*)-type magnetic Bragg peaks and no anomaly has been found at $T^*$ [14]. As a result, the feature at $T^*$ in magnetic susceptibility measurements is ascribed to a magnetic domain process [14]. In contrast, a distinct magnetic easy axis has been proposed more recently, where the magnetic moments are determined to be along the *c* axis since the (0 0 *L*)-type magnetic Bragg peaks are found absent. In addition, a kink-like feature has been reported in the temperature dependence of the Bragg reflection (2 0 0) at $T^*$, which is also claimed to be observed in all other reflections [15]. Therefore, the easy axis of the magnetic moments and the nature of the anomaly at $T^*$ of $Sr_4Ru_3O_{10}$ remain elusive.

Another intriguing property of $Sr_4Ru_3O_{10}$ is the field-induced metamagnetic transition. While a typical magnetic hysteresis loop of ferromagnets is observed as the magnetic field is applied along the *c* axis, a first-order metamagnetic transition is seen below $T^*$ for the field applied in the *ab* plane [10,11]. Experimental signatures of the electronic phase separation during the phase transition have been observed [16]. Nevertheless, previous neutron diffraction studies reveal controversial features regarding this metamagnetic transition. Across the critical field, while no anomaly has been observed in the field dependence of the magnetic reflection (0 0 8) in Ref. [14], a field-induced transition at the same wave vector is reported in Ref. [15]. In addition, it has been reported that the metamagnetic transition is accompanied by a change in the lattice, as evidenced in Raman scattering [13], neutron diffraction [15], and magnetostriction studies [17], which suggests the presence of strong spin-lattice coupling in this system. To date, the nature and the origin of this metamagnetic transition are yet to be resolved.

In this paper, we present a single-crystal neutron diffraction study of the magnetic order and the metamagnetic transition in $Sr_4Ru_3O_{10}$. We find that the material orders ferromagnetically at $T_c \approx 100$ K without any signature of antiferromagnetic components, in agreement with the previous studies [14,15]. However, the magnetic moments are found to possess both in-plane and out-of-plane components below $T_c$, in contrast to the previous neutron diffraction studies where the spins are proposed to align either in the *ab* plane [14] or along the *c* axis [15]. In addition, we have observed spin reorientations below $T^* \approx 50$ K, where the magnetic moments incline toward the out-of-plane direction. Furthermore, we show that while the magnetic moments are nearly along the *c* axis at $T = 1.5$ K, a magnetic field applied along the in-plane [1 −1 0] direction gives rise to a spin reorientation at a critical field $B_c$, above which there exists a spin component perpendicular to the plane defined by the field direction and the *c* axis, in accordance to the metamagnetic transition. This observation is distinct from the previous studies, where the metamagnetic transition is ascribed to magnetic domain processes [14] or a coexistence of zero-field *c*-axis component and the field-induced in-plane spin polarization [15]. Both temperature- and magnetic-field-driven spin reorientations are presumably ascribable to the change in the magnetocrystalline anisotropy that is strongly correlated with the $RuO_6$ octahedral distortion. Our results reconcile the inconsistency among previous neutron diffraction studies on the magnetic easy axis of $Sr_4Ru_3O_{10}$, resolve the puzzling character of the transition at $T^*$, and elucidate the nature of the field-induced metamagnetic transition.

**Results**

We start with the zero-field neutron diffraction data collected at the HB-3A four-circle diffractometer. Figure 2(a) and 2(b) present the rocking curves of Bragg reflections (0 0 2) and (1 1 1) at $T$ = 8, 50 and 100 K, respectively. Note that both reflections meet the requirements of nuclear Bragg diffraction. The intensity of (0 0 2) shows a nonmonotonic behavior, which is significantly enhanced at 50 K compared to that measured at 8 K and 100 K; in contrast, the (1 1 1) intensity becomes stronger with decreasing temperature. These observations suggest that there is magnetic intensity superimposed on the nuclear Bragg peaks. To further elucidate the nonmonotonic temperature dependence of the magnetic intensity, Figure 2(c) and 2(d) present the peak intensity of (0 0 2) and (1 1 1) Bragg reflections as a function of temperature, respectively. Upon cooling, additional reflection intensity in both peaks emerges below 100 K. Similar enhancement has been observed in other nuclear Bragg peaks including (0 0 6) and (0 0 8), indicating the development of a ferromagnetic order below $T_c$ = 100 K which is in line with the magnetic susceptibility measurements [10]. Consistent with Fig. 2(a), the intensity of (0 0 $L$)-type magnetic reflections, $L$ = 2, 6 and 8, also exhibits nonmonotonic temperature dependence, with a maximum at $T^* \approx$ 50 K that is reminiscent of the anomaly at $T^*$ observed in the magnetic susceptibility data [10,11]. In accordance to Fig. 2(b), the magnetic intensity at (1 1 1) emerges at $T_c$ and continues to increase below $T^*$, which suggests a spin reorientation at $T^*$. On the one hand, the observation of (0 0 $L$)-type magnetic reflections is in agreement with the previous neutron diffraction experiments reported in Ref. [14], but is different from that in Ref. [15] where they are found to be absent. On the other hand, we have revealed distinct temperature-dependent behaviors of the intensity of (0 0 $L$) and (1 1 1), which is in sharp contrast to the previous studies where either no anomaly was observed [14], or a kink feature was claimed to exist in the temperature dependence of all reflections [15]. Note that the HB-3A diffractometer is equipped with a two-dimensional neutron detector, from which the change in the scattering angle 2θ of the diffracted neutrons at different temperatures is found too small to be resolved due to the instrumental resolution limitation. This is in agreement with a change in the lattice constants of only ~0.04% upon cooling observed in the previous studies [15]. Therefore, the variation in the peak intensity of these Bragg peaks cannot be ascribed to the structural change, but is magnetic in origin.

In order to explore the possible magnetic configurations in $Sr_4Ru_3O_{10}$, we carried out the magnetic representational analysis using the program SARAh [18]. The space group of the crystal structure is No. 55 *Pbam* and the propagation vector is **k** = (0 0 0). There are four inequivalent Ru atoms in a chemical unit cell which are located at Ru1 (0 0 0), Ru2 (0 0 0.1402), Ru3 (0.5 0 0.3598) and Ru4 (0.5 0 0.5). The analysis shows that an in-plane antiferromagnetic component would give rise to strong magnetic reflections at (1 0 0), (1 0 1) or (0 1 1), which, nevertheless, are all absent in our measurements. Therefore, the feature at $T^*$ in the magnetic susceptibility cannot be ascribed to the formation of antiferromagnetic components in the *ab* plane that was claimed previously [11,13]. As a result, the only symmetry-allowed magnetic configurations in $Sr_4Ru_3O_{10}$ is the ferromagnetic structure with the magnetic moments either along the *c* axis or in the *ab* plane. Since neutrons couple to the magnetic moment perpendicular to the momentum transfer **q**, the observation of magnetic intensities at (0 0 $L$), $L$ = 2, 6, and 8, at $T^* < T < T_c$ suggests the development of a magnetic order with spins aligned in the *ab* plane. This, combined with the magnetic susceptibility measurements which show an easy axis closer to the *c* axis, suggests that the magnetic moments have both in-plane and out-of-plane components in this temperature regime. Note that at $T^* < T < T_c$, the intensity of (0 0 2) increases more rapidly than that of (1 1 1), which implies that the magnetic moments incline toward the *ab* plane, though the easy axis is always closer to the $c$ axis. Below $T^*$, the decrease in the intensity of (0 0 $L$)-type magnetic peaks suggests that the in-plane ferromagnetic moment reduces, and the spins incline continuously toward the out-of-plane direction upon cooling. This speculation is supported by the observation of an increase in the reflection intensity of the (1 1 1) Bragg peak below $T^*$, as the direction of the (1 1 1) wave vector is almost perpendicular to the $c$ axis (note that $a^* \approx b^* \approx$ 1.137 Å$^{-1}$, $c^* \approx$ 0.219 Å$^{-1}$). We have collected the nuclear and magnetic Bragg peaks at 8 K, 50 K and 100 K, and attempted to perform Rietveld analysis using the program FullProf [19]. Note that the magnetic moments of the outer-layer Ru ions (Ru1 and Ru4) and the central-layer ones (Ru2 and Ru3) can be different, owing to the difference in the direction and magnitude of the RuO$_6$ octahedral rotation [10]. At $T$ = 8 K, due to the very weak intensity of the (0 0 $L$) peaks, the spins are nearly aligned along the $c$ axis. Since only the (1 1 1) peak is associated with this spin component, the magnitude of the magnetic moments of outer- and central-layer Ru ions cannot be exclusively determined at this temperature. Similarly, more magnetic reflections are required in order to determine the moment sizes and the canting angles quantitatively at $T^*$. The schematic of the magnetic structure at zero field is plotted in Fig. 1(b).

Next, we discuss the neutron diffraction measurements done on the HB-1A triple-axis spectrometer with the magnetic field applied along the in-plane [1 -1 0] direction, where a first-order metamagnetic transition is observed in the magnetic susceptibility measurements [10]. Figure 3(a) and 3(b) display the peak intensity of (0 0 2) and (1 1 1) Bragg peaks as a function of magnetic field at $T$ = 1.5 K. It is remarkable that the (0 0 2) Bragg peak intensity increases significantly at $B_c$ = 1.5 T, whereas that of the (1 1 1) decreases. These observations are distinct from the previous studies where no anomaly has been observed in the intensity of magnetic reflections as a function of magnetic field [14]. It is worth noting that in another neutron diffraction study a similar trend to (0 0 2) [Fig. 3(a)] has been observed in the field dependence of the intensity of (0 0 8) Bragg peak [15]. However, a slight reduction in the scattering intensity of (H K 0) reflections was claimed and suggested to be associated with the magnetic moment along the c axis, which led the authors to argue that the field-induced transition is due to a coexistence of the initial zero-field ferromagnetic order with a field-induced in-plane spin polarization [15]. If this were the case, one would expect a slight increase in the magnetic intensity of (1 1 1) reflection above the critical field $B_c$, as (1 1 1) is perpendicular to [1 -1 0] but is about ~82.2° with respect to the [0 0 1] direction. This is not in line with our observation of the field-induced decrease in the intensity of (1 1 1) reflection, which suggests that the applied magnetic field gives rise to a spin component perpendicular to the plane defined by the field direction and the c axis. This is supported by a recent study which, by measuring the magnetic moment vectors, reports the existence of a magnetic moment component that is perpendicular to the field rotation plane [20]. Note that the slight difference in the critical field compared with the magnetic susceptibility data may arise from different applied magnetic field directions in the ab plane. To further support the variation of the magnetic scattering intensity below and above $B_c$, Figure 3(c) and 3(d) plot the θ-2θ scans over (0 0 2) and (1 1 1) Bragg peaks measured at T = 1.5 K with B = 0 and 3.5 T, respectively. We can clearly see that above $B_c$ the intensity of (0 0 2) peak becomes enhanced while that of the (1 1 1) peak is suppressed. Furthermore, there is no noticeable change in the 2θ values of both peak positions across the field-induced magnetic transition, which is consistent with the previous study that the lattice constants a and c change only by ~0.04% at the critical field that is beyond the instrumental resolution in our study [15]. For comparison, in the inset of Fig. 3(d), we present the θ-2θ scan of a high-|Q| nuclear peak (0 0 16), where the magnetic intensity is expected to be very weak due to the small magnetic form

factor. The very little difference in the intensity of (0 0 16) at 0 and 3.5 T magnetic field further substantiates that the field-induced intensity variation at low-|**Q**| peaks (0 0 2) and (1 1 1) are magnetic in origin, rather than a structural change. Figure 4(a) and 4(b) show the θ-2θ scans of (0 0 2) and (1 1 1) Bragg peaks measured at $T$ = 50 K with $B$ = 0 and 3.5 T, respectively. Both (1 1 1) and (0 0 2) scattering intensity displays negligible changes upon applying the magnetic field. These results, combined with the features observed at low temperature shown in Figure 3, are in agreement with the magnetic susceptibility measurements that the metamagnetic transition only occurs below $T^*$ = 50 K.

**Discussions**

The observation of a spin reorientation at $T^*$ in $Sr_4Ru_3O_{10}$ at zero field has elucidated the long-standing puzzle on the anomaly in the magnetic susceptibility measurements [10]. This spin reorientation suggests a change in the magnetic easy axis as a function of temperature. This feature is reminiscent of the widely studied spin reorientation transitions in rare earth magnets and orthoferrites [21], which are ascribed to the change in the magnetic anisotropy constants as a function of external parameters, such as temperature, magnetic field and pressure, etc [22]. Our finding raises an intriguing question: What is the underlying mechanism responsible for the change in the magnetic easy axis in $Sr_4Ru_3O_{10}$? It is known that in Ruddlesden-Popper type ruthenates, the magnetic anisotropy is strongly coupled to the structural distortions of the $RuO_6$ octahedra. For instance, magnetocrystalline anisotropy has been extensively investigated in $SrRuO_3$ thin films, where the magnetic anisotropy and the saturated magnetic moments can be readily tuned by controlling the $RuO_6$ octahedral distortion via the epitaxial strain imposed by substrates [23]. Considering the fact that the ferromagnetic state with $T_c \approx 100$ K in triple-layer ($n$ = 3) $Sr_4Ru_3O_{10}$ bridges the physics of the double-layer ($n$ = 2) $Sr_3Ru_2O_7$ (Fermi liquid, ferromagnetic instability) [2] and the three-dimensional ($n = \infty$) SrRuO$_3$ (ferromagnetic, $T_c$ = 160 K) [24], the change in the magnetocrystalline anisotropy in Sr$_4$Ru$_3$O$_{10}$ across $T^*$ may arise from the corresponding changes in lattice structures. Indeed, anomalies in the lattice constants with $c$ expanding while $a$ contracting below $T^*$ at zero field have been observed previously [15,17]. These changes in lattice parameters are expected to alter the occupancy of the Ru $t_{2g}$ orbitals in this multiband system, which consequently affects the magnetic anisotropy through spin-orbit coupling [25].

The nature of the in-plane field-induced metamagnetic transition in Sr$_4$Ru$_3$O$_{10}$ below $T^*$, which has been an unresolved puzzle, can now be readily understood. It is also due to a spin reorientation where the magnetic moments reorient from the nearly $c$ axis toward the $ab$ plane. Intriguingly, we find that there exists a magnetic moment component perpendicular to the plane defined by the field direction and the $c$ axis above the critical field, which is consistent with the observation of a recent study where it has been ascribed to the Dzyaloshinskii-Moriya interaction assuming that the easy axis is along the $c$ axis [20]. On the other hand, it is worth noting that across the field-induced metamagnetic transition below $T^*$ the lattice parameter $a$ expands while $c$ shrinks [15], a trend which is opposite to that observed upon cooling through $T^*$ at zero field discussed above [15,17]. This suggests that the magnetic anisotropy is altered above the critical field due to the strong spin-lattice coupling, which needs to be taken into account in the future theoretical modeling.

In summary, the magnetic structure and the metamagnetic transition in the triple-layer ruthenate Sr$_4$Ru$_3$O$_{10}$ ($n = 3$) are revisited by neutron diffraction measurements. The magnetic order below $T_c$ is found to be ferromagnetic with both in-plane and out-of-plane spin components, while it evolves toward the $c$ axis below $T^* = 50$ K. This finding elucidates the long-standing puzzle of the nature of the magnetic transition at $T^*$ in $Sr_4Ru_3O_{10}$. Below $T^*$, upon the metamagnetic transition with a magnetic field applied along the in-plane [1 -1 0] direction, the magnetic moments undergo a spin reorientation from the $c$ axis toward the $ab$ plane with a spin component perpendicular to the plane defined by the magnetic field direction and the $c$ axis. Both temperature- and field-induced spin reorientations can be attributed to the change of magnetocrystalline anisotropy via spin-orbit coupling due to the change in the $RuO_6$ octahedral distortion.

**Methods**

The single-crystal $Sr_4Ru_3O_{10}$ was grown by the floating zone method [26]. The lattice constants are $a$ = 5.5280 Å, $b$ = 5.5260 Å and $c$ = 28.651 Å. The phase and quality of the crystal have been verified by *x*-ray diffraction measurements. The magnetization as a function of both temperature and magnetic field has been measured using SQUID, and the results are in agreement with previous studies [10, 11]. A very small amount of intergrowth of $SrRuO_3$ has been revealed in the magnetic susceptibility measurements, which is easy to be separated in the neutron diffraction measurements due to its very different lattice parameter $c$ ($c$ = 7.8446 Å in orthorhombic unit cell). Zero-field and field-dependent neutron diffraction measurements were performed using the HB-3A four-circle diffractometer ($\lambda$ = 1.5426 Å) and the HB-1A triple-axis spectrometer ($\lambda$ = 2.36 Å) respectively at High Flux Isotope Reactor in Oak Ridge National Laboratory. At HB-3A, the sample was mounted on an aluminum stick and loaded into a closed-cycle Helium displex. At HB-1A, the sample was oriented in the horizontal (*H H L*) scattering plane and loaded into a vertical-field cryomagnet such that the magnetic field is applied along the [1 -1 0] direction. The Bragg reflections (*H K L*) were in the reciprocal lattice units $2\pi/a$, $2\pi/b$ and $2\pi/c$ of the orthorhombic space group *Pbam* (No. 55) [10].

**Data availability**

The data that support the findings of this study are available from the corresponding author upon reasonable request.

**Acknowledgements**

Work at Michigan State University was supported by the National Science Foundation under Award No. DMR-1608752 and the start-up funds from Michigan State University. Work at Tulane University was supported by the U.S. Department of Energy (DOE) under EPSCOR Grant No. DE-SC0012432 with additional support from the Louisiana Board of Regents (support for crystal growth). A portion of this research used resources at the High Flux Isotope Reactor, a DOE Office of Science User Facility operated by the Oak Ridge National Laboratory.

**Author Contributions**

X. K. conceived the project. P. G. L., Y. W., and Z. Q. M. grew the single-crystal sample. M. Z., H. B. C., W. T., X. K., H. D. Z., and B. D. P. performed neutron diffraction experiments. M. Z., H. B. C., W. T., and X. K. analyzed the data. M. Z. and X. K. wrote the manuscript. All authors commented on the manuscript.

**Additional information**

Competing Interest: The authors declare that they have no competing financial interests.
15

**Figure Captions**

Fig. 1. (a) The crystal structure of $Sr_4Ru_3O_{10}$. Sr, Ru and O atoms are represented by the green, gray and red balls, respectively. (b) The magnetic structure of $Sr_4Ru_3O_{10}$. Only the magnetic Ru ions are shown. Note that both moment size and spin direction cannot be uniquely determined due to the lack of enough Bragg peaks with reasonably good magnetic intensity in the measurements.

Fig. 2. (a),(b) Rocking curve scans across (0 0 2) and (1 1 1) Bragg reflections at $T$ = 8, 50 and 100 K, respectively. (c),(d) Temperature dependence of the peak intensity of (0 0 2) and (1 1 1) reflections at $B$ = 0 T, respectively. The red and green dashed lines denote $T_c$ and $T^*$, respectively.

Fig. 3. (a),(b) Field dependence of the peak intensity of (0 0 2) and (1 1 1) reflections at $T$ = 1.5 K. (c),(d) θ-2θ scans across (0 0 2) and (1 1 1) Bragg reflections at $B$ = 0 and 3.5 T, $T$ = 1.5 K. Inset shows the θ-2θ scan over (0 0 16) peak at $B$ = 0 and 3.5 T, $T$ = 1.5 K.

Fig. 4. θ-2θ scans across (0 0 2) and (1 1 1) Bragg reflections at $B$ = 0 and 3.5 T, $T$ = 50 K.

**Figure 1**

M. Zhu et al.

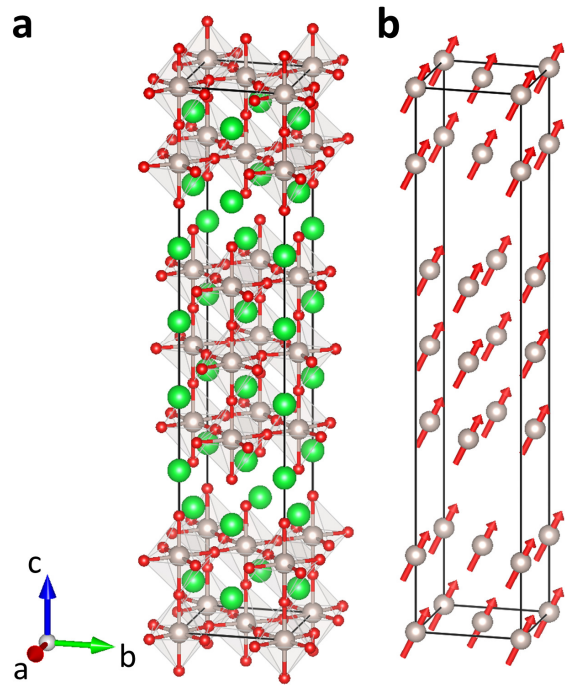

**Figure 2**

M. Zhu et al.

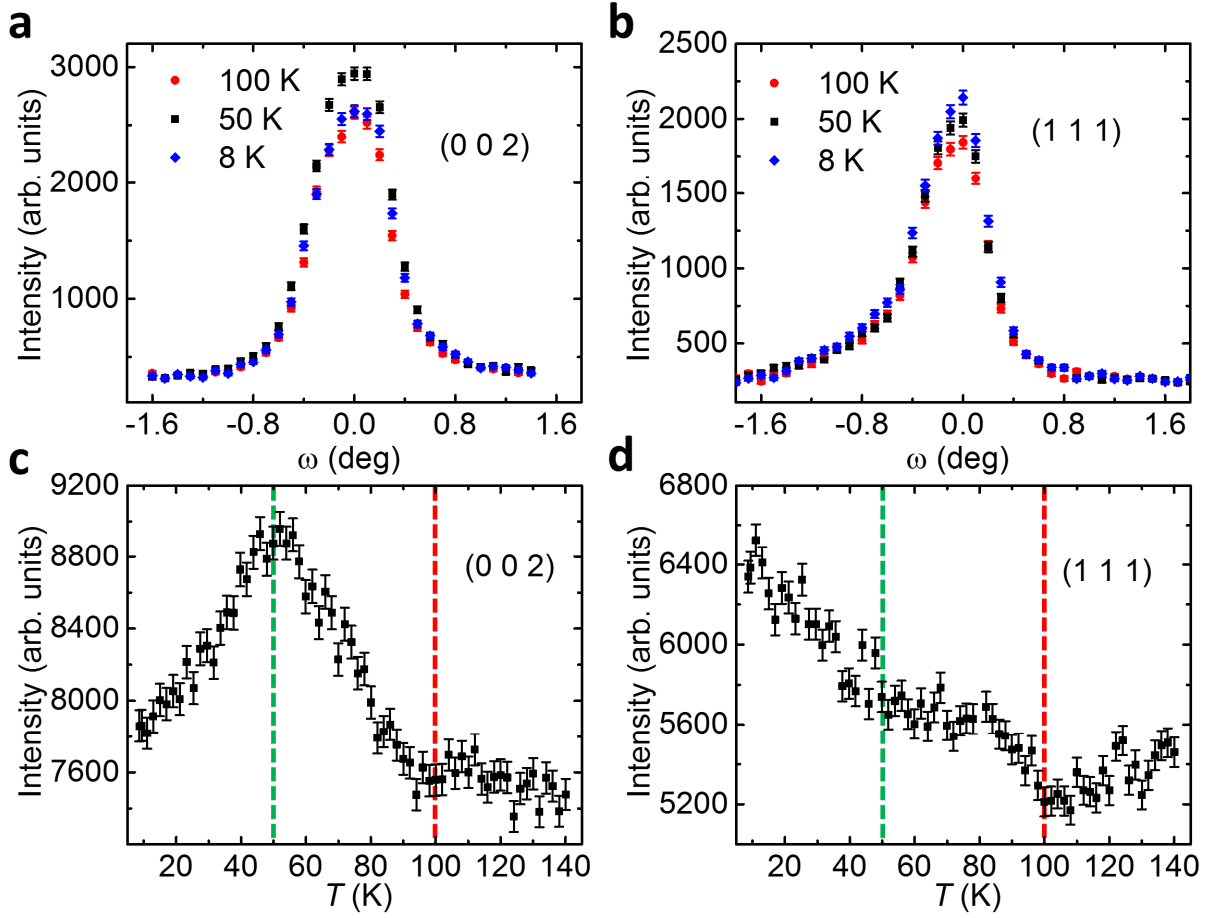

**Figure 3**

M. Zhu et al.

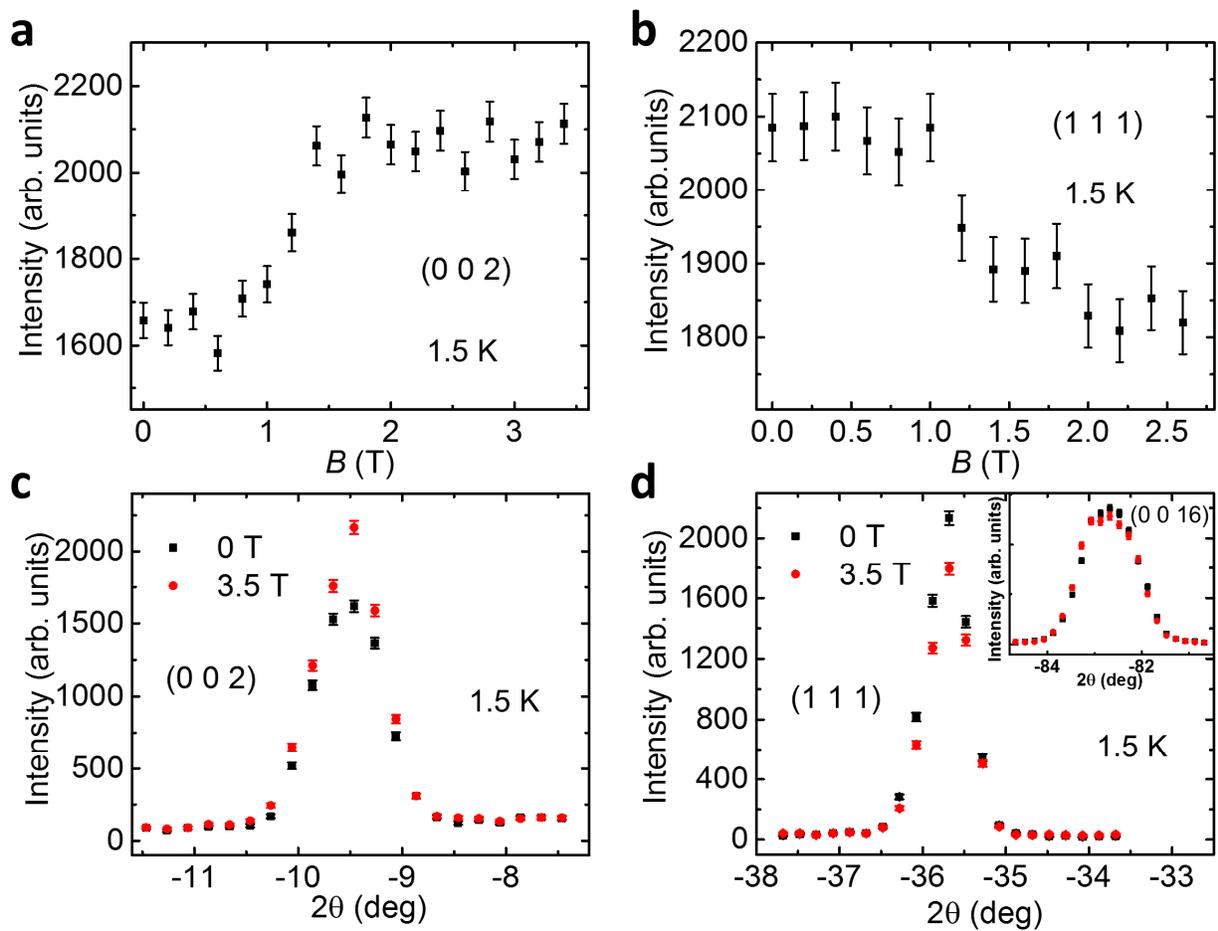

**Figure 4**

M. Zhu et al.

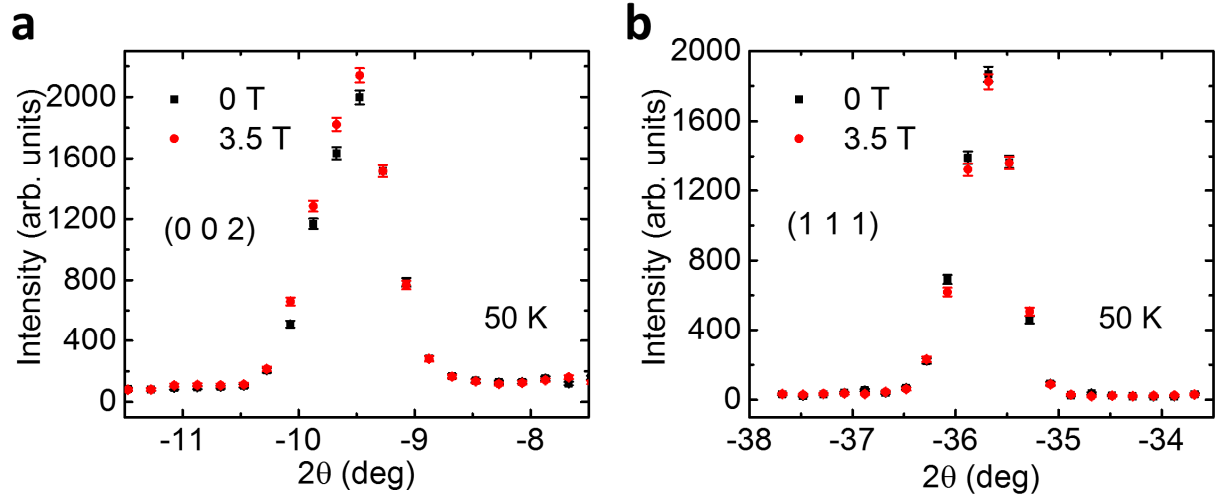